\documentclass[12pt]{article}
\usepackage{fancyhdr}
\usepackage{url}
\usepackage{hyperref}

 {\fancyhead{}
\fancyfoot[c]{\small{\rule{17.5cm}{1pt}}}}

\begin{document}
\title{\vspace{-2.5cm}
\begin{center}
\textbf{\small{ECIR WORKSHOP REPORT}}\\\vspace{-0.5cm} \rule{17.5cm}{1pt}
\end{center}
\vspace{1cm}\textbf{Report on TBAS 2012: Workshop on Task-Based and Aggregated Search}}

\author{Birger Larsen \\
       Royal School of Library \\ 
       and Information Science\\
       Denmark\\
       \emph{blar@iva.dk}
       \and
       Christina Lioma \\
       University of Copenhagen\\
        Denmark\\
       \emph{liomca@gmail.com} \\
       \and
       Arjen de Vries \\
       Centrum Wiskunde \\
       and Informatica\\
       The Netherlands\\
       \emph{arjen@acm.org} \\     
       \date{07 April 2012}}

\maketitle \thispagestyle{fancy} \abstract{
The ECIR half-day workshop on Task-Based and Aggregated Search (TBAS) was held in Barcelona, Spain on 1 April 2012. The program included a keynote talk by Professor J\"{a}rvelin, six full paper presentations, two poster presentations, and an interactive discussion among the approximately 25 participants. This report overviews the aims and contents of the workshop and outlines the major outcomes.


\section{Introduction}
TBAS 2012\footnote{\url{http://itlab.dbit.dk/~tbas2012/}} was a half-day workshop organised as part of the European Conference on Information Retrieval (ECIR) on 1 April 2012, in Barcelona, Spain. TBAS aimed to investigate task-based and aggregated search and to practically stimulate exploratory research in these areas.

\textbf{Task-based search} aims to understand the user's current task and desired outcomes, and how this may provide useful context for the Information Retrieval (IR) process \cite{IngwersenJ:2005}. An example of task-based search is situations where a query given by a user is too vague or ambiguous to produce good retrieval results. In such cases, additional information from the user on e.g. the purpose of the search or what the user already knows about the topic can provide valuable additional evidence that can significantly improve retrieval performance \cite{KellyF07}. For example, in an academic search setting this could include references to relevant earlier papers or names of pertinent researchers in the field. Task-based search may be especially useful in cases of \textbf{aggregated search} \cite{ArguelloDCC11}, also known as integrated search \cite{stone:2010} in the digital libraries domain. Aggregated search describes the increasingly common IR paradigm of presenting in a single search box one result list with information from different document and media types, such as Wikipedia entries, Webpages, user-authored content, images, locations, etc. Research into aggregated search addresses the challenge of when and how to fuse different document and media types, and how to present results to the user. An example of aggregated search is the retrieval of scientific content, which involves searching among different domain-dependent document types and structures (e.g. full articles, short abstracts, tables of content). In this scenario, knowledge of the user's task and problem at hand may be beneficial to the aggregated search process. 

Task-based and aggregated search are two challenging IR tasks, with special research opportunities and possibilities. However, they are not well supported by current test collections in IR. The driving motivation of this workshop was to stimulate exploratory research in these two areas, by offering a test collection (iSearch \cite{LykkeLLI10}) that has been designed specifically to support task-based and aggregated retrieval research. The collection was made available to workshop participants to allow them to investigate perspectives, to present results, to facilitate and stimulate discussion, and to engage researchers with novel ideas and viewpoints who might not have the means to create benchmark test collections for assessing aspects of task-based and aggregated retrieval.  

\section{The iSearch Test Collection}
iSearch\footnote{\url{http://itlab.dbit.dk/~isearch/}} (7GB) is a test collection of scientific documents from the physics domain, collected from arXiv.org and from the union catalogue for all Danish libraries (in English). 

In terms of task-based search, iSearch comes with 65 topics and their relevance assessments, created by 23 lecturers and experienced graduate students from 3 different university departments of physics. The topics represent real information seeking tasks. Each topic contains 5 fields: (1) description of the information sought, (2) user background, (3) work task, (4) ideal answer, (5) keywords. The relevance assessment of each topic was made by the same user who formulated that topic, by examining a pool of documents retrieved for that topic. Relevance was assessed on a 4-point scale: highly, fairly, marginally, and non-relevant. 

In terms of aggregated search, iSearch comprises: (i) full length articles, (ii) article metadata records, (iii) book metadata records. The book records include almost all books in English on physics published in the period 1999-2009 that are to be found in Danish research and university libraries, in the British National Bibliography, and in the Library of Congress in USA. In addition, iSearch includes 3.7 million internal citations to facilitate research in citation and network analysis for bibliometrics and IR.

The iSearch test collection, which has so far been used for research in IR \cite{LiomaKS11,Lykke:2010} and NLP \cite{MichelbacherKFLS11}, is the only aggregated search test collection whose relevance assessments were explicitly designed to cover different types of documents, e.g. books and articles, for the same topics.

\section{Workshop Program}
The workshop started with an invited keynote talk by Professor Kalervo J\"{a}rvelin, aimed to outline the problems and form a joint understanding for workshop attendees of the issues and challenges of task-based and aggregated search. In the main session, six full papers and two posters were presented\footnote{The presentations are available at the TBAS 2012 website: \url{http://itlab.dbit.dk/~tbas2012/}. The proceedings are available permanently from \url{http://ceur-ws.org/}}. The workshop finished with a lively interactive discussion among the participants. Overall, approximately 25 people attended the workshop. 

\subsection{Keynote}
Professor Kalervo J\"{a}rvelin (University of Tampere, Finland) talked about \textit{Information Access and Integration in Task-based Heterogeneous Environments}.
The talk drew on experiences from an extensive study on task-based information access of molecular medicine researchers, their needs to integrate multiple (types of) information resources, and the problems they encountered in using them. Methods for integrating heterogeneous (XML) resources were examined: identifying potentially relevant resources, examining them for their usefulness, and solving problems of inconsistency of data/document structures, labels, and values. 
The talk concluded that there is considerable interaction between systems in real world use, and that these systems should not be designed (or evaluated) in isolation, and that information access may benefit from aggregated search, integration of resources by linking, and from provision of suitable search keys and data harmonisation.

\subsection{Full Paper Presentations}
Out of the six full papers accepted, four mainly focused on task-based search and two on aggregated search. 
\subsubsection{Task-based Search} 
Grzywaczewski et al. \cite{Adam} discussed the development of task-specific IR systems for software engineers. Based on a user study consisting of questionnaires and automated observation of user interactions with the browser and software development environment, they first discussed how software engineers interact with information and IR systems. A notable result was that the behaviour between users was very similar and stable over time, and that it mainly consisted of reuse of existing code when available and highly relevant. The authors then investigated to what extent a domain-specific search and recommendation system can be developed in order to support their work related activities. Finally, they discussed factors that can be used as implicit feedback indicators for further collaborative filtering and discussed how these parameters can be analysed using computational intelligence based techniques.

Beckers and Fuhr \cite{Beckers} presented a conceptual model for the user-oriented design of IR systems supporting different kinds of search tasks. Based on a new architectural model for IR systems, they showed how the different levels of such a system should be modified in order to provide better support for task-based search. In particular they argued that the functional level must be enriched, by considering cognitive models and implementing appropriate functionalities. A main point was that the system interface cannot compensate for missing system functionalities.

Ingwersen and Wang \cite{Peter} analysed the relationship between usefulness assessments and perceptions of work task complexity and search topic specificity as provided by the authors of the 65 topics of the iSearch test collection. It was found that highly specific topics associate with all degrees of task complexity, whereas highly complex tasks tend to be associated with search topics of high specificity. 

Norozi et al. \cite{Norozi} presented work on using citation information as contextual evidence for boosting relevant documents. The authors proposed a re-weighting model to contextualise bibliographic evidence in a query-independent and query-dependent fashion (based on Markovian random walks). The inlinks and outlinks of a node in the citation graph were used as context, based on the hypothesis that documents in a good context (having strong contextual evidence) should be relevant with respect to a query. The proposed models were experimentally evaluated using the iSearch Collection and found to significantly outperform state of the art baselines.

\subsubsection{Aggregated Search} 
S\o rensen et al. \cite{Sorensen} studied the usability of various known retrieval techniques for aggregated search. These techniques included optimising indexing settings (stemming, stopword removal) and retrieval parameters, biasing the ranking according to document weights specific to document types (e.g. boosting the ranking of books as opposed to metadata), as well as merging the ranked lists of documents using various fusion approaches. Their experiments with the iSearch collection concluded that aggregated search across different document types in the digital library domain can benefit by optimising and boosting the search process separately for each type of document.

Lu et al. \cite{Lu} addressed the problems of displaying aggregated results to the user, so that different types of documents are fairly represented in a single ranked list. They studied two methods of maximising the number of relevant results of each type of aggregated document that are displayed to the user. The first method used the well-known clarity score query performance predictor and the second method used the difference in retrieval status value between the first ranked document and documents at different cut-offs further down the ranked list. These methods were evaluated in a simulated setting: instead of asking users directly which display they prefer, they investigated which settings maximise the number of relevant documents displayed to the user. They found query performance prediction to be useful in that task, a finding that paves the way for further research in the use of query perfromance prediction for optimally displaying aggregated search results.

\subsection{Poster Presentations}

Sawitzki et al. \cite{Frank} presented an information portal for the domain of social sciences that supports key aspects of aggregated search in a user-friendly
way. Such aspects include metadata information, links, as well as visualisation of the aggregated search results in meaningful and non-intrusive ways.

Dragusin et al. \cite{Radu} presented a task-based search engine for the medical domain, and specifically for the sub-domain of rare disease retrieval by clinicians. They presented two functionalities of their system: the visual clustering of search results by disease name, and the optional ranking of search results by disease relevance, as opposed to document relevance.

\subsection{Interactive Discussion}
Throughout the workshop, an interactive discussion between workshop participants took place. Peter Ingwersen acted as devil's advocate in the background, and participants in the interactive discussion were, among others, Kalervo J\"{a}rvelin, Mounia Lalmas, Peiling Wang, Udo Kruschwitz, Gene Golovchinsky, Thomas Beckers, Adam Grzywaczewski, Radu Dragusin, and Adam Zhou.

A first topic was the dilemma between on one hand offering a large number of functionalities (that might aid specific tasks and empower the user) and on the other keeping the interface sufficiently simple (as to not overload the user). Some systems have a very large number of functionalities and expect the user to selectively personalise search tools and sessions by learning only those functionalities needed for given tasks. Some dedicated users may be able to do this but many may be overloaded or simply ignore the advanced functionalities. It was discussed how to then identify which tools to present in which contexts, with the conclusion that this is a very challenging task that requires a high level of understanding of the user context and task. This can be very hard to achieve in general domains, but examples were given for more limited domains where we are now seeing progress in specialised applications that perform quite well. 

A second topic discussed was that often multiple systems are used in combination to solve any given task, and that these are most often developed independently by different developers without an understanding of the wider use in different contexts. It was concluded that there is great scope in developing systems that are more homogeneous and have greater understanding of each other. 

Other topics discussed was that there are interesting links to other emerging IR perspectives, such as the session-based retrieval and the focus here is on search missions encompassing many different searches in different systems. It was also remarked that most of the presented work focussed either on task-based or aggregated search, but that none really tackled both perspectives. Overall, the consensus emerging from the interactive discussion is that a task-based perspective has great potentials for aggregated search and that in reality aggregation takes place implicitly. Doing research in combining the two perspectives is challenging and there is still a long way to go, but there are signs that research is moving in this direction.

\subsection{Best Paper/Poster Awards}
As part of the reviewing process of the TBAS submissions, the program committee was asked to nominate and rate best papers and posters. According to these ratings, the best paper award was given to Beckers and Fuhr for their work \textit{Towards the Systematic Design of IR Systems Supporting Complex Search Tasks} \cite{Beckers}. The best poster award was given Sawitzki et al. for their work \textit{Extending Aggregated Search in a Social Sciences Digital Library} \cite{Frank}. We congratulate the authors for their work. 

\section{Conclusions and Future Directions}
The TBAS 2012 workshop featured an excellent keynote and a wide range of interesting papers and posters from both academia and industry. Several facets of task-based and aggregated search were presented and discussed in a highly interactive and lively environment. The workshop was successful in bringing people from these research communities and from industry together, and in underlying the potential for synergies between task-based and aggregated search. A follow-up workshop next year is under consideration.

\paragraph{Acknowledgements}
The TBAS organisers wish to thank BCS-IRSG and ECIR for hosting the workshop. The organisers also wish to thank the ECIR workshop chair, David Losada, the TBAS 2012 keynote speaker, Prof. Kalervo J\"{a}rvelin (University of Tampere, Finland), and the TBAS program committee for their valuable contributions and feedback: Jaime Arguello (University of North Carolina at Chapel Hill, USA), Pablo Castells (Autonomous University of Madrid, Spain), Miles Efron (University of Illinois at Urbana-Champaign, USA), Nicola Ferro (University of Padova, Italy), Norbert Fuhr (University of Duisburg-Essen, Germany), Hideo Joho (University of Tsukuba, Japan), Jaap Kamps (University of Amsterdam, The Netherlands), Diane Kelly (University of North Carolina at Chapel Hill, USA), Mounia Lalmas (Yahoo! Research Barcelona, Spain), Monica Landoni (University of Lugano, Switzerland), Marie-Francine Moens (Katholieke Universiteit Leuven, Belgium), Ian Ruthven (University of Strathclyde, UK), Tassos Tombros (Queen Mary University of London, UK), Elaine Toms (University of Sheffield, UK), Pertti Vakkari (University of Tampere, Finland). Finally, special thanks go to all the authors, presenters and participants of TBAS 2012 for their contributions and thought-provoking discussions that formed a succesfull workshop.

\bibliographystyle{abbrv}
\bibliography{larsenLV}

\begin{thebibliography}{10}

\bibitem{ArguelloDCC11}
J.~Arguello, F.~Diaz, J.~Callan, and B.~Carterette.
\newblock A methodology for evaluating aggregated search results.
\newblock In P.~Clough, C.~Foley, C.~Gurrin, G.~J.~F. Jones, W.~Kraaij, H.~Lee,
  and V.~Murdock, editors, {\em ECIR}, volume 6611 of {\em Lecture Notes in
  Computer Science}, pages 141--152. Springer, 2011.

\bibitem{Beckers}
T.~Beckers and N.~Fuhr.
\newblock Towards the systematic design of ir systems supporting complex search
  tasks.
\newblock In Larsen et~al. \cite{tbas2012}, pages 29--33.

\bibitem{Radu}
R.~Dragusin, P.~Petcu, C.~Lioma, and O.~Winther.
\newblock Zebra: Searching for rare diseases a case of task-based search in the
  medical domain.
\newblock In Larsen et~al. \cite{tbas2012}, pages 36--39.

\bibitem{Adam}
A.~Grzywaczewski, R.~Iqbal, J.~Halloran, K.~Iqbal, and A.~James.
\newblock Copy and paste as an indicator of relevance: Towards the development
  of intelligent recommender systems for software engineers.
\newblock In Larsen et~al. \cite{tbas2012}, pages 14--18.

\bibitem{IngwersenJ:2005}
P.~Ingwersen and K.~J\"{a}rvelin.
\newblock {\em The Turn: Integration of Information Seeking and Retrieval in
  Context (The Information Retrieval Series)}.
\newblock Springer-Verlag New York, Inc., Secaucus, NJ, USA, 2005.

\bibitem{Peter}
P.~Ingwersen and P.~Wang.
\newblock Relationship between usefulness assessments and perceptions of work
  task complexity and search topic specificity: An exploratory study.
\newblock In Larsen et~al. \cite{tbas2012}, pages 19--23.

\bibitem{KellyF07}
D.~Kelly and X.~Fu.
\newblock Eliciting better information need descriptions from users of
  information search systems.
\newblock {\em Inf. Process. Manage.}, 43(1):30--46, 2007.

\bibitem{tbas2012}
B.~Larsen, C.~Lioma, and A.~P. de~Vries, editors.
\newblock {\em Proceedings of the Task-Based and Aggregated Search (TBAS2012)
  Workshop, ECIR 2012, Barcelona, Spain, 01-April-2012}.
  \url{http://ceur-ws.org}, 2012.

\bibitem{LiomaKS11}
C.~Lioma, A.~Kothari, and H.~Sch{\"u}tze.
\newblock Sense discrimination for physics retrieval.
\newblock In W.-Y. Ma, J.-Y. Nie, R.~A. Baeza-Yates, T.-S. Chua, and W.~B.
  Croft, editors, {\em SIGIR}, pages 1101--1102. ACM, 2011.

\bibitem{Lu}
W.~Lu, Q.~Wang, and B.~Larsen.
\newblock Simulating aggregated interfaces.
\newblock In Larsen et~al. \cite{tbas2012}, pages 24--28.

\bibitem{Lykke:2010}
M.~Lykke, P.~Ingwersen, T.~Bogers, H.~Lund, and B.~Larsen.
\newblock Physicists' information tasks: structure, length and retrieval
  performance.
\newblock In {\em Proceedings of the third symposium on Information interaction
  in context}, IIiX '10, pages 347--352, New York, NY, USA, 2010. ACM.

\bibitem{LykkeLLI10}
M.~Lykke, B.~Larsen, H.~Lund, and P.~Ingwersen.
\newblock Developing a test collection for the evaluation of integrated search.
\newblock In C.~Gurrin, Y.~He, G.~Kazai, U.~Kruschwitz, S.~Little, T.~Roelleke,
  S.~M. R{\"u}ger, and K.~van Rijsbergen, editors, {\em ECIR}, volume 5993 of
  {\em Lecture Notes in Computer Science}, pages 627--630. Springer, 2010.

\bibitem{MichelbacherKFLS11}
L.~Michelbacher, A.~Kothari, M.~Forst, C.~Lioma, and H.~Sch{\"u}tze.
\newblock A cascaded classification approach to semantic head recognition.
\newblock In {\em EMNLP}, pages 793--803. ACL, 2011.

\bibitem{Norozi}
M.~A. Norozi, A.~P. de~Vries, and P.~Arvola.
\newblock Contextualization from the bibliographic structure.
\newblock In Larsen et~al. \cite{tbas2012}, pages 9--13.

\bibitem{Frank}
F.~Sawitzki, P.~Schaer, and D.~Hienert.
\newblock Extending aggregated search in a social sciences digital library.
\newblock In Larsen et~al. \cite{tbas2012}, pages 34--35.

\bibitem{Sorensen}
D.~R. S$\o$rensen, T.~Bogers, and B.~Larsen.
\newblock An exploration of retrieval-enhancing methods for integrated search
  in a digital library.
\newblock In Larsen et~al. \cite{tbas2012}, pages 4--8.

\bibitem{stone:2010}
G.~Stone.
\newblock Searching life, the universe and everything? {T}he implementation of
  {Summon at the University of Huddersfield}.
\newblock {\em Library Quarterly}, 20(1):24--52, 2010.

\end{thebibliography}
\end{document}